\documentclass[conference]{IEEEtran}
\usepackage{amsmath,graphicx}
\graphicspath{{images/}}
\usepackage{amsfonts}

\ifCLASSINFOpdf
\else
\fi
\begin{document}
\title{Twitter User Geolocation using Deep Multiview Learning}

\author{
\IEEEauthorblockN{Tien Huu Do, Duc Minh Nguyen, Evaggelia Tsiligianni, Bruno Cornelis, Nikos Deligiannis}
\IEEEauthorblockA{Vrije Universiteit Brussel, Pleinlaan 2, B-1050 Brussels, Belgium \\
                          imec, Kapeldreef 75, B-3001 Leuven, Belgium \\
                          \{thdo, mdnguyen, etsiligi, bcorneli, ndeligia\}@etrovub.be}
}


%Telephone: (800) 555--1212, Fax: (888) 555--1212}


\maketitle

\begin{abstract}
Predicting the geographical location of users on social networks like Twitter is an active research topic 
with plenty of methods proposed so far. 
Most of the existing work follows either a content-based or a network-based approach. 
The former is based on user-generated content while the latter exploits the structure of the network of users. 
In this paper, we propose a more generic approach, which incorporates not only both content-based and network-based features, 
but also other available information into a unified model. 
Our approach, named Multi-Entry Neural Network (MENET), leverages the latest advances in deep learning and multiview learning. 
A realization of MENET with textual, network and metadata features results in an effective method for Twitter user geolocation, 
achieving the state of the art on two well-known datasets.
\end{abstract}
%
%\begin{keywords}
%Twitter user geolocation, multiview learning, deep learning, feature learning.
%\end{keywords}
%
%
%
%----------------------------------------------------%
\section{Introduction}
\label{sec:intro}
%----------------------------------------------------%
Social networks have become more and more popular, with billions of active users on a daily basis.
Among the most widely-used social networks, Twitter stands out as an attractive option, with a unique mechanism of 
publishing short messages, termed \textit{tweets} and re-posting messages, termed \textit{retweets}. 
This way information can be broadcasted widely and quickly through the Twitter network. 
As one of the most popular social networks, a lot of useful, yet unstructured, information is available on Twitter. 
User location is essential for a wide range of applications such as social unrest forecasting~\cite{compton2013detecting}, event detection~\cite{guille2014mention} and location-based 
service recommendation~\cite{bao2015recommendations}. 
Nevertheless, the availability of geo-tagged tweets and geolocation-enabled user profiles 
on Twitter is highly limited \cite{ICWSM1510561}. As a result, automatically analysing and predicting user's location from 
Twitter data is of great significance, and has received a lot of attention from both industry and academia. 

The task of predicting users' locations on Twitter is often referred to as the \textit{Twitter User Geolocation} problem. 
Several algorithms have been proposed so far to solve this problem. 
Existing algorithms can be categorized into two broad groups, namely content-based and network-based approaches. 
While content-based algorithms \cite{ICWSM1510561,LiuPaper} exploit 
textual contents from tweets, network-based algorithms \cite{ICWSM136067,compton2014geotagging} 
make use of the connections and interactions between users for the task of predicting user's location. 
Both approaches have achieved good location accuracy until now \cite{ICWSM1510561,RahimiCB17}.

%This paper focuses on a more generic approach for the Twitter user geolocation problem. 
%Leveraging recent advances in deep neural networks~\cite{bengio2015deep, schmidhuber2015deep} and multiview learning \cite{ZHAO201743},
%we propose a model that combines the strong points from both groups of Twitter user geolocation algorithms. 

This paper focuses on a more generic approach for the Twitter user geolocation problem by leveraging recent advances in deep neural networks~\cite{bengio2015deep} and multiview learning~\cite{ZHAO201743}. Deep neural networks have been proven to be very effective in many domains including image classification~\cite{krizhevsky2012imagenet}, image super-resolution~\cite{zhou2017guided}, speech recognition~\cite{graves2013speech} and compressive sensing~\cite{nguyen17}. On the other hand, multiview learning, which considers learning with multiple feature sets to improve the generalization performance, has made a great progress recently~\cite{zhang2012combining,yu2012combining}.
Based on these techniques, our model effectively predicts users' locations from Twitter data and 
achieves state-of-the-art results in well-known benchmarks. 
Our contribution in this paper is three-fold:
\begin{itemize}
   \item We propose a neural network architecture, named multi-entry neural network (MENET), for Twitter user geolocation.
   MENET is capable of combining multiview features into a unified model to infer users' locations.
   \item We propose a realization of MENET in a Twitter user geolocation method with four specific types of features.
   \item We present an extensive experimental evaluation on popular benchmarks. The experiments show that our method achieves state-of-the-art results.
\end{itemize}

The remainder of this paper is organized as follows. Section~\ref{sec:related_work} briefly describes related work. 
Section~\ref{sec:architecture} explains our method in detail, including the model architecture, feature extraction and learning. 
Section~\ref{sec:experiments} presents our experimental settings and results. 
Finally, we conclude our paper in Section~\ref{conclusion_future_work}.
%
%
%
%----------------------------------------------------%
\section{Related Work}\label{sec:related_work}
%----------------------------------------------------%
Two main approaches were proposed in the literature for the Twitter user geolocation problem. 
The first approach, which has been investigated thoroughly, uses textual features from tweets for building location predictive models. 
The second approach, on the other hand, arises from the observation that a user often interacts with people in the vicinity, 
and exploits the network connections of users. 
This section brings a closer look on recent works in both approaches.

%Plenty of content-based methods have been proposed for Twitter user geolocation. 
%Geographic topic models~\cite{hong2012discovering,Eisenstein:2010:LVM:1870658.1870782} 
%consider tweets and locations as the outputs of a generative process incorporating topics and regions as latent variables. 
%Gaussian Mixture Model is later employed to model the relation between geographic areas and terms of tweets ~\cite{priedhorsky2014inferring, Chang:2012:PTP:2456719.2457115}.
%These methods, however, do not take into account the distribution of datasets over the regions of interest. Addressing this problem, grid-based geolocation is introduced~\cite{Roller:2012:STG:2390948.2391120,Wing:2011:SSD:2002472.2002593,Wing_hierarchicaldiscriminative,melo2015geocoding}. For these methods, an adaptive or uniform grid is created to partition the datasets into cells at different levels. The prediction of geographic coordinates then is converted to a classification problem using cells as classes, and off-the-shelf classifiers can be applied directly. This strategy is also employed in this paper where the administrative boundaries of geographic areas are taken into account. Not using a grid, Liu \& Inkpen~\cite{LiuPaper} train stacked denoising autoencoders for predicting regions and states, and for predicting coordinates. Char \textit{et al.}~\cite{ICWSM1510561} estimate location exploiting the expressiveness of sparse coding and the advance in dictionary learning to obtain the state of the art on a benchmark dataset named GeoText compared to other content-based methods.

Plenty of \textit{content-based} methods have been proposed for Twitter user geolocation using 
geographical topic models~\cite{eisenstein2010}
and Gaussian Mixture Models (GMM)~\cite{priedhorsky2014inferring}.
More recently, Liu and Inkpen~\cite{LiuPaper} trained stacked denoising autoencoders for predicting location of Twitter users. 
Char \textit{et al.}~\cite{ICWSM1510561} estimated the location by exploiting the expressiveness of sparse coding to obtain state-of-the-art results on a benchmark dataset named GeoText~\cite{eisenstein2010}.
These methods, however, do not take into account the distribution of users' locations over the regions of interest. 
Addressing this problem, grid-based geolocation is introduced in 
~\cite{roller2012,Wing_hierarchicaldiscriminative},
where an adaptive or uniform grid is created to partition the datasets into appropriate cells. 
The prediction of geographic coordinates then is converted to a classification problem using cells as classes. 

The key idea behind the \textit{network-based} approach is that 
there is a correlation between the likelihood of friendship of two social network users 
and their geographical distance~\cite{backstrom2010find}. 
Using this correlation, the location of a user can be revealed via his or her friends' location. 
By leveraging social interactions like bi-directional following~\cite{journals/tgis/DavisPOA11} and bi-directional mentioning~\cite{compton2014geotagging}, 
one can establish a graph where label propagation~\cite{raghavan2007near} or its variants are used 
to identify location of unlabeled users. 
The weakness of this method is that it can not propagate labels (locations) to users who are not connected to the graph. 
To address this problem, methods combining textual information and graph topology knowledge are proposed in~\cite{RahimiCB15,RahimiCB17}. Furthermore, these works build densely undirected graphs based on mentioning of users, 
which helps improve significantly the results. 
A similar mention graph is utilized in this paper. However, instead of using label propagation directly on the graph like in~\cite{RahimiCB15,RahimiCB17}, we rely on an efficient embedding to capture the graph structure.
%
%
%
%----------------------------------------------------%
\section{Multiview Learning Architecture}
\label{sec:architecture}
%----------------------------------------------------%
%% problem statement 
In Twitter user geolocation,
we consider a Twitter user of interest and collect multiple tweets over a period of time. 
Each tweet contains not only textual content, but also metadata information such as the posting timestamp. 
Semantic analysis of tweets can reveal information about the location of the user.
Textual data can also be used to retrieve information concerning the user's interaction with other users,
i.e., to create a network of users.
Together with metadata, these types of information consist the multiple views of the model proposed in the present work.

The task of Twitter geolocation is to predict the location of a user 
in terms of geographical region or exact geocoordinates (longitude and lattitude).
Predicting the geographical region is a classification problem.
Exact prediction of geocoordinates is a regression problem;
however, here, we also address this problem as a classification problem. 
Every region is assigned a pair of geocoordinates 
corresponding to the median value (centroid) of the geocoordinates of all the sample users belonging to that region.
After classifying a new user to a region,
we use the region's centroid as an estimation of the user's location. 

We propose a generic neural network model, 
referred to as Multi-Entry Neural Network model (MENET),
to leverage multiple views of the Twitter data for this task.
We realize our generic model into an effective Twitter user geolocation system, with four 
different types of features.
Next, we present the different feature types employed in our realization, 
followed by the proposed MENET architecture. 

%------------------------------------------------------------%
\subsection{Multiview Features}
%------------------------------------------------------------%

%The employment of different types of features aims at the fusion of different types of information.
A common way to employ unstructured information into machine learning models is the use of embeddings to bring the  information into a structured form.
We build representations of textual information using word and paragraph embeddings
such as Term Frequency - Inverse Document Frequency (TF-IDF)~\cite{MassiveDataMining} and doc2vec~\cite{DBLP:journals/corr/LeM14}.
The user network information is represented using an algorithm known as node2vec~\cite{grover2016node2vec}.
In our realization, we have also employed a timestamp feature to leverage information concerning the posting time of tweets
which is often related with user's location.
Next, we describe the four feature types employed in the proposed MENET architecture. 

\subsubsection*{TF-IDF Feature}
Term Frequency - Inverse Document Frequency (TF-IDF)~\cite{MassiveDataMining} is a statistical measure used to evaluate how important a term is to a document in a corpus. 
The importance of a term increases proportionally to the number of times a term appears in the document (TF) but is offset by the frequency of the term across the corpus (IDF). 
TF-IDF is then defined as a product of TF and IDF values. 
In this work, we consider a document a concatenation of tweets posted from the same user. 
We employ the well-implemented library \textit{scikit-learn}~\cite{scikit-learn} to 
calculate TF-IDF feature from the document
%In this work, we employ the well-implemented library \textit{scikit-learn}~\cite{scikit-learn} to 
%calculate TF-IDF feature from the document, which created by concatenating tweets from the same user. 
%
%
\subsubsection*{Context Feature}
The context feature is a mapping from a variable-length document,  
to a fixed-sized continuous valued vector. 
This vector provides a numerical representation, capturing the context of the document. 
Originally proposed in~\cite{DBLP:journals/corr/LeM14},
the context feature is also referred to as doc2vec or \textit{Distributed Representation of Sentences},
and it is an extension of the broadly used \textit{word2vec} model~\cite{DBLP:journals/corr/MikolovSCCD13}.

In this work, we employ the Distributed Bag of Words of Paragraph Vector algorithm (PV-DBOW)~\cite{DBLP:journals/corr/LeM14} 
for extracting the doc2vec feature from the Twitter documents. 
The PV-DBOW algorithm delivers robust performance if trained on large datasets \cite{DBLP:journals/corr/LeM14}. 
Our implementation utilizes the Gensim library \cite{rehurek_lrec} for both training and extracting features.

\subsubsection*{Node2vec Feature}\label{node2vec_feature}
Whereas the TF-IDF and doc2vec features capture textual content of the tweets, 
with the third feature, we aim to capture the information of a Twitter user's network. 
In particular, from the given tweets, we build a user network graph with each node corresponding to a user, 
and employ the node2vec algorithm~\cite{grover2016node2vec} 
to extract continuous feature representations for each node. 
Given a set $V$ of nodes, the basic idea of node2vec is to learn a function 
$f: V \rightarrow \mathbb{R}^d$ that maps each node into a $d$-dimensional feature space which 
preserves the connectivity patterns of the whole network. 

Our user graph is formed in a way similar as in~\cite{RahimiCB17},~\cite{RahimiCB15} but instead of predicting users' locations directly on the graph, we extract node2vec feature for later use in our model.
First, a unique set of nodes, $V$, is created for all the users of interest. 
If a user mentions directly another user and both of them belong to $V$, an edge is created reflecting this interaction. 
The weight of an edge is equal to the number of mentions between the two corresponding users. 
Moreover, if two users of interest mention a third user, who may or may not belong to $V$, 
we create an edge between these two users, with a weight equal to the sum of the mentioning times. 
In addition, we define celebrity users as users with a high number of unique connections with regard to a pre-defined threshold. 
We remove all connections to these celebrities since celebrities often have a huge number of active connections, thus mentioning a celebrity is much less likely to reveal geographical relation. 
%This graph building procedure is depicted in Fig.~\ref{fig:graph}.
%
%\begin{figure}[h]
%\centering
%\includegraphics[scale=0.45]{graph}
%\centering
%\caption{Twitter user graph via mentioning. User\_1 mentions User\_2, thus, we create an edge between them. User\_1 and User\_3 are connected because they both mention External User.}
%\label{fig:graph}
%\end{figure}

\subsubsection*{Timestamp Feature}
In many Twitter databases like GeoText~\cite{eisenstein2010} and UTGeo2011~\cite{roller2012}, 
the posting time (timestamp) of all tweets is available in terms of the Coordinated Universal Time (UTC). 
In~\cite{twiter_geo_location_timing_matters}, it was shown that there exists a relation between the time and location of a Twitter stream. 
In fact, it is less likely that people tweet late at night than at any other time, which implies a drift in longitudes. 
Therefore, the variation in timestamp could be an indication for longitude. 
We obtain the timestamp feature for a given user as follows.
First, we extract the timestamps from all the tweets of that user
and convert them to the standard format to extract the hour.
After that, a $24$-dimensional vector is created corresponding to $24$ hours in a day;
the $i$-th element of this vector equals the number of messages posted by the user at the  $i$-th hour. 
Finally, this feature vector is $\ell_2$ normalized.

\subsection{Model Architecture}
\label{Model_Architecture}
\begin{figure}[t]
\centering
\includegraphics[width=0.9\linewidth]{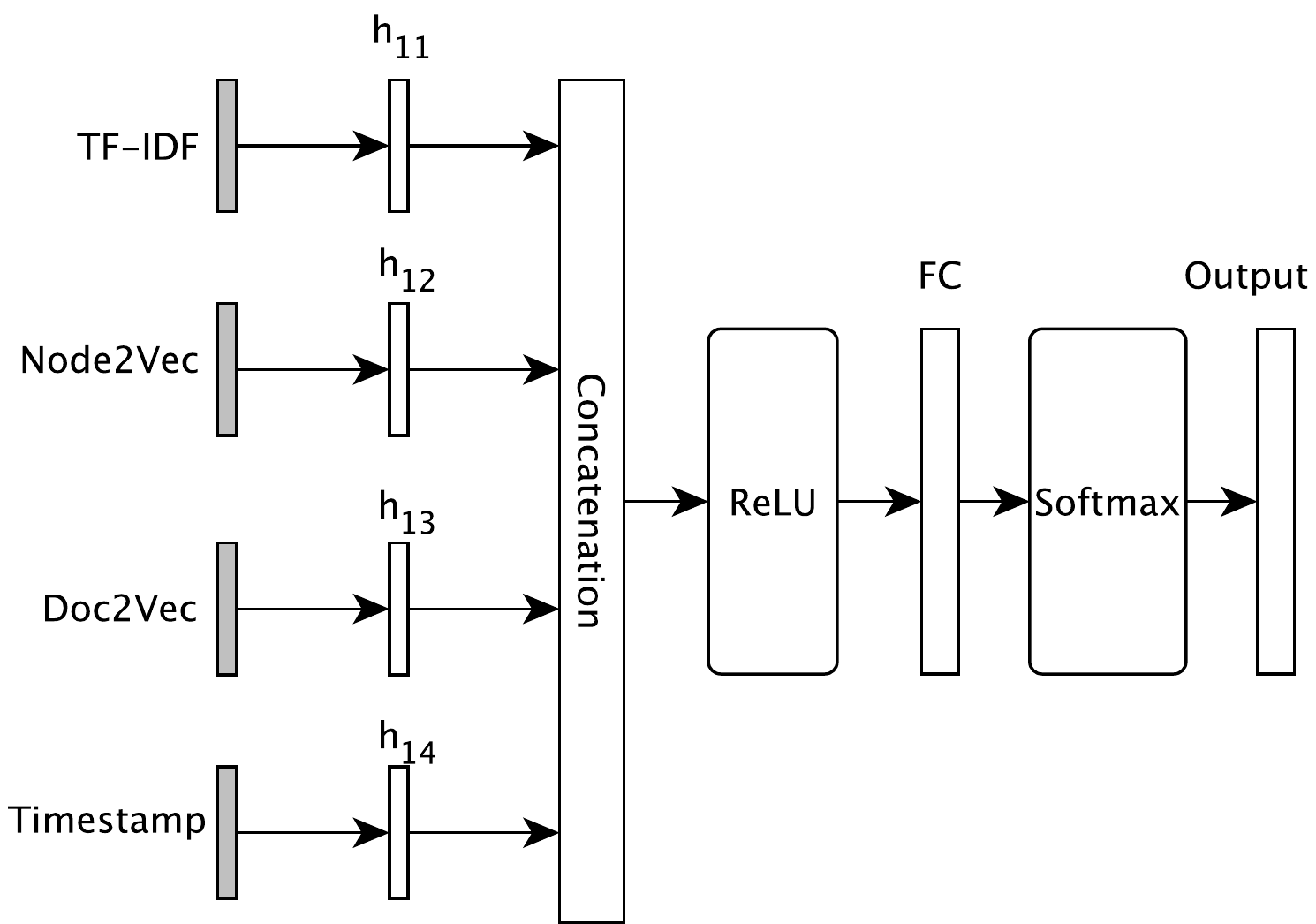}
\centering
\caption{The proposed multi-entry neural network architecture}
\label{fig:model1}
\end{figure}

%\subsubsection*{Architecture}
%
%The proposed model is illustrated in Fig.~\ref{fig:model1}. 
%We refer to it as Multi-Entry Neural Network model, or MENET for short. 
The proposed model is illustrated in Fig.~\ref{fig:model1}.
The model takes as input different types of feature vectors. 
Each feature vector corresponds to one view, capturing specific information of the Twitter data. 
Using different views of the available Twitter data, 
the model classifies the respective user into one of the predefined classes
corresponding to geographical regions. 
With four feature types employed, our realization of MENET in this work 
has four views, as shown in Fig.~\ref{fig:model1}.

Each view is the input to one branch in MENET, which is one fully connected hidden layer. 
The hidden layer is followed by a Rectified Linear unit (ReLU)~\cite{article:relu} activation function. 
Each branch realizes a non-linear function,  
mapping the original feature vectors into a unified feature space. 
In the learned feature space, the outputs from all views are concatenated to form a compact representation of each user.  
This representation serves as the input to a classifier with one fully-connected (FC) layer. 
This classifier employs an $m$-way softmax to transform 
scores into class probabilities, with $m$ the number of classes. 

It is worth mentioning that a more straight-forward approach to combine multiple features is to concatenate them 
before inputting into the network. Nevertheless, we argue that our architecture is more effective. 
The simple concatenation of the original feature vectors results in input vectors of high dimensions, 
which significantly increase the number of parameters in the model and make the model more prone to overfitting. 
Although the number of hidden units in each layer is adjustable, we opt to 
set the output of each branch, except for the timestamp, to be of lower dimension than their corresponding inputs. 
As a result, each branch can be seen as a dimensionality reduction function. 
This way, we can mitigate the overfitting problem during training.  

We formulate the Twitter user geolocation task as a classification problem, and 
employ the cross-entropy loss as the objective function to train our model.
Considering $m$ classes of users, the cross-entropy loss over $n$ training samples is given by 
\begin{equation} \label{eq:cross_entropy}
	L = -\sum_{i=1}^{n} \sum_{j=1}^{m} y_{i}^{j} \log(\tilde{y}_i^{j}), 
\end{equation}
where $y_{i}$, $i=1,\dots , n$, is the ground-truth vector for sample $i$, 
$\tilde{y}_i$ is the vector holding the predicted probabilities of sample $i$ for each class,
and $y_{i}^{j}$ denotes the $j$-th element of the respective vector.
%

%\subsubsection*{Training and Testing MENET}\label{overfitting_handling}
We train our MENET model using the Stochastic Gradient Descent (SGD) algorithm. 
In order to control overfitting, we employ a weight decay regularizer and early stopping strategy. 
Particularly, during training, the model performance in a seperated validation set is monitored. 
If this performance decreases for a pre-defined number of epochs, the training is stopped. 
We also anneal the learning rate as the training proceeds. 
	
At the testing stage, we compare the predicted location classes with the ground truth labels 
to measure the model's performance in terms of accuracy. 
\section{Experiments}\label{sec:experiments}
\subsection{Datasets}
In order to evaluate our proposed model, we employ two datasets, namely, 
the GeoText \cite{eisenstein2010} and UTGeo2011~\cite{roller2012} datasets. 
The GeoText dataset contains approximately $370$K tweets from $9475$ users in the US, collected during the first week of March, 2010. 
This dataset is splitted into non-overlapping subsets of $7580$, $1895$ and $1895$ users, for training, validation and testing, respectively. 
Compared to the GeoText dataset, the UTGeo2011 dataset is larger, with $38$M tweets collected from $449$K users in the US. 
In this dataset, $429$K users, approximately, are reserved for training, while each one of the validation and testing sets consists of $10$K users. 
In both datases, all tweets from a specific user are concatenated to form a single document. 
Following~\cite{ICWSM1510561,RahimiCB15,RahimiCB17}, the location prediction is performed at user level, with the ground-truth location of each user defined as the geocoordinates of the their first tweet. 
The location is characterized by two numbers, the longitude and latitude values. 
%We create the state labels from the ground-truth user coordinates
%using the Ray Casting algorithm~\cite{shimrat1962algorithm}.

\begin{table}[t] 
\centering
\caption{Hyperparameter setting for MENET. $n_{h_{11}}, n_{h_{12}}, n_{h_{13}}, n_{h_{14}}$ denote the number of neurons
 in the hidden layers $h_{11}, h_{12}, h_{13}, h_{14}$ for the features TF-IDF, node2vec, doc2vec and timestamp, respectively.}
\label{table:regional_classification_hyperparameter}
%\resizebox{\columnwidth}{!}{%
 \begin{tabular}{| c | c |}
 \hline
  & Number of hidden units  \\
 \hline
  \hline
  $n_{h_{11}}$ & 150   \\ 
 \hline
  $n_{h_{12}}$  & 150  \\
 \hline
 $n_{h_{13}}$ & 30   \\
 \hline
 $n_{h_{14}}$ & 30   \\
 \hline
\end{tabular}
%}
\end{table}

\begin{table}[t]
\centering
\caption{Regional and state classification accuracy results on the GeoText and UTGeo2011 datasets. N/A stands for not available.}
\label{table:classification_result}
%\resizebox{\columnwidth}{!}{
 \begin{tabular}{|c | c c | c c|} 
 \hline
  & \multicolumn{2}{|c|}{GeoText} & \multicolumn{2}{|c|}{UTGeo2011} \\
  & Region  & State & Region  & State \\
  & (\%) & (\%) & (\%) & (\%) \\
\hline
 \hline
Eisenstein \textit{et al.}~\cite{eisenstein2010} & 58 & 27 & N/A & N/A \\
\hline
Liu \& Inkpen~\cite{LiuPaper} & 61.1 & 34.8 & N/A & N/A \\
\hline
Cha \textit{et al.}~\cite{ICWSM1510561} & 67 & 41 & N/A & N/A \\
\hline
MENET & \textbf{76} & \textbf{64.8} & 83.7 & 69 \\ 
\hline
\end{tabular}
%}
\end{table}

\begin{table*}[t!]
\centering
\caption{Performance comparison on geographical coordinates prediction. The results include the mean and median distance errors and the accuracy within 161 kilometers.  
N/A stands for not available.}
\label{table:coordinates_prediction}
 \begin{tabular}{| c | c c c | c c c | c c c |}
 \hline
  & \multicolumn{3}{|c|}{GeoText} & \multicolumn{3}{|c|}{UTGeo2011}  \\ [0.5ex] 
 & mean & median & @161 & mean & median & @161 \\
  & (km) & (km) &(\%) &(km) &(km) &(\%) \\
 \hline
\hline
Eisenstein \textit{et al.}~\cite{eisenstein2010} & 900 & 494 & N/A & N/A & N/A & N/A \\
%\hline
%Wing \textit{et al.}~(2011)~\cite{Wing:2011:SSD:2002472.2002593} & 967 & 479 & N/A & N/A & N/A & N/A\\ 
 \hline
Roller \textit{et al.}~\cite{roller2012}  & 897 &432 & 35.9 & 860 & 463 & 34.6 \\
 \hline
%Wing \& Baldridge~(Uniform)~\cite{Wing_hierarchicaldiscriminative} & N/A & N/A & N/A & 703.6 & 170.5 & 49.2 \\
% \hline
%Wing \& Baldridge~(KD tree)~\cite{Wing_hierarchicaldiscriminative} & N/A & N/A & N/A & 686.6 & 191.4 & 48.0 \\
%  \hline
%Melo \textit{et al.}~\cite{melo2015geocoding} & N/A & N/A & N/A & 702 & 208 & N/A \\
% \hline
Liu and Inkpen~\cite{LiuPaper} & 855.9 & N/A & N/A & 733 & 377 & 24.2 \\
 \hline
Cha \textit{et al.}~\cite{ICWSM1510561} & 581 & 425 & N/A & N/A & N/A & N/A  \\
 \hline
Rahimi \textit{et al.}~(2015)~\cite{RahimiCB15} & 581 & \textbf{57} & 59 & 529 & 78 & 60 \\
 \hline
Rahimi \textit{et al.}~(2017)~\cite{RahimiCB17} & 578 & 61 & 59 & 515 & \textbf{77} & \textbf{61} \\ 
 \hline
  MENET & \textbf{570} & 58 & \textbf{59.1} & \textbf{474} & 157 & 50.5 \\ 
 \hline
\end{tabular}
\end{table*}

\subsection{Performance Criteria}
We evaluate our model in three different tasks: 
(i) four-way classification of US regions including Northeast, Midwest, West and South; 
(ii) fifty-way classification at US state level; 
(iii) estimation of the real-valued user coordinates. 
For the first two tasks, we report the classification accuracy, 
whereas, for the latter task, we report the mean and median distance errors. 
We also compare our model against reference methods in terms of 
the accuracy measure \textit{@$161$}, which is defined as the percentage of predictions  
with a distance error less than $161$ km\footnote{$161$~km $\sim$ $100$ mile}.
All distance measures between coordinates are computed using the Haversine formula~\cite{1984S.T.68R.158S}. 
We compare the results of our models to those of recent reference methods in 
\cite{eisenstein2010}, 
\cite{ICWSM1510561}, \cite{roller2012}, 
\cite{RahimiCB15}
and \cite{RahimiCB17}. 

%--------------------------------------------
\subsection{Implementation Details}
%--------------------------------------------
We implement our model using Tensorflow\footnote{https://www.tensorflow.org/}. 
We first pre-process the Twitter data by performing tokenization, removing stop words, URLs and punctuation, 
and finally stemming the words. These pre-processing steps are implemented using the NLTK library~\cite{Bird:2009:NLP:1717171}. 

Concerning the features, TF-IDF features are extracted using the \textit{scikit-learn} library \cite{scikit-learn}, with 
minimum term frequency across documents set to $40$ and $500$ for the GeoText and UTGeo2011 datasets, respectively. 
We utilize the original source codes provided by the authors in~\cite{grover2016node2vec}, \cite{DBLP:journals/corr/LeM14} to extract the node2vec 
and doc2vec features and set both feature types to have $300$ dimensions. 

In our experiments, we empirically configure our model for each dataset. The model's configuration is shown in Table~\ref{table:regional_classification_hyperparameter}. During training, we use a small learning rate, $\alpha=0,0001$ 
and regularize the weight of the output layer, with the regularization parameter $\lambda$ set to $0.1$. 
The training procedure is done using the ADAM optimization algorithm ~\cite{DBLP:journals/corr/KingmaB14}.

\subsection{Results}

The regional and state classification results are shown in Table \ref{table:classification_result}. 
As can be seen from this table, our model significantly outperforms the state of the art in both region and state levels on the 
GeoText dataset. By leveraging the classification strength of multiple features, 
the improvement in regional accuracy is $9\%$ compared to the state-of-the-art results presented in~\cite{ICWSM1510561}. 
Concerning the state classification, the achieved accuracy is $64.8\%$ compared to $41\%$ in  \cite{ICWSM1510561}.
It should be noted that the classification results of the reference methods 
on the UTGeo2011 dataset are not available. 

The comparison between all methods on the geocoordinate prediction task is presented in Table \ref{table:coordinates_prediction}. 
Our MENET model performs overall the best on the GeoText dataset. 
Compared to \cite{RahimiCB17}, our model has lower mean  
and median distance errors, and marginally higher accuracy measure \textit{@$161$}. 
On the UTGeo2011 dataset, our MENET model also outperforms all reference methods 
in terms of mean distance error. Nevertheless, \cite{RahimiCB15} and \cite{RahimiCB17} 
achieve better results for other metrics. 
It should be noted that these two methods 
employ map partitioning strategies to create new classes optimized for each dataset taking into account the geographical distribution of users. Requiring each class to have approximately the same number of users, the partitioning algorithm yields more balanced classes: high density regions (classes) are divided into smaller areas resulting in better accuracy. Large areas results in lower prediction accuracy. 
On the other hand, our method relies on the administrative boundaries of regions and states, ignoring the users's distribution.
This brings an adverse effect on our method.
%This makes the median distance error and the accuracy @161 not as good as in~\cite{RahimiCB15,RahimiCB17}. 
However, the map partitioning strategies are independent from the network architecture and can be applied to the proposed MENET model.  
We leave this exploration for our future work. 

%It should be noted that these two methods 
%employ map partitioning strategies to create new classes optimized for each dataset taking into account the geographical distribution of users. This leads to more balanced classes, namely a class with plenty of users has a small area and vice versa. On the other hand, our method uses the administrative boundaries of regions and states, ignoring the users's distribution. Some states, for example Montana, have very little users while their areas are very large. This makes the median distance error and the accuracy measure @161 not as good as in~\cite{RahimiCB15,RahimiCB17}.
%However, the map partitioning strategies are independent from the network architecture and can be applied to our MENET model. 
%We leave this exploration for our future work. 

%-----------------------------------------------------------%
\section{Conclusion}
\label{conclusion_future_work}
%-----------------------------------------------------------%
Twitter user geolocation is a challenging task because of insufficient labelled training data. 
The linguistically noisy nature of the Twitter data and the excessive size of the Twitter network make the task even harder. 
While there exist several approaches in the literature, the problem of attaining a high accuracy still remains open.
In this paper, we follow the multiview learning paradigm by combining knowledge from both user-generated content and network interaction.
In particular, we propose a neural network model, referred to as MENET, 
that uses word frequency, paragraph semantics, network topology and timestamp information, 
to infer users' locations. Our model achieves state-of-the-art results and can be easily extended to leverage other types of information, besides the considered types of data.

%Our future work will focus on investigating better geographical partitioning schemes to improve the prediction of geo-coordinates and extending the geolocation problem to wider areas including the world.

%In our future work, we will focus on making the model truly inductive, meaning able to generalize to new users. 
%Inspired from recently published results on graph evolving strategies, 
%our research directions include the development of algorithms that produce robust embeddings for unseen nodes,
%which will be applied to infer geolocation for new Twitter users.

%\vfill
%\pagebreak

%\vfill\pagebreak
%\newpage

\bibliographystyle{IEEEbib}
\bibliography{refs}

\begin{thebibliography}{10}

\bibitem{compton2013detecting}
R.~Compton, C.~Lee, T.~Lu, L.~D. Silva, and M.~Macy,
\newblock ``Detecting future social unrest in unprocessed twitter data:
  emerging phenomena and big data,''
\newblock in {\em IEEE International Conference on Intelligence and Security
  Informatics}, 2013, pp. 56--60.

\bibitem{guille2014mention}
A.~Guille and C.~Favre,
\newblock ``Mention-anomaly-based event detection and tracking in twitter,''
\newblock in {\em IEEE/ACM International Conference on Advances in Social
  Networks Analysis and Mining}, 2014, pp. 375--382.

\bibitem{bao2015recommendations}
J.~Bao, Y.~Zheng, D.~Wilkie, and M.~Mokbel,
\newblock ``Recommendations in location-based social networks: a survey,''
\newblock {\em GeoInformatica}, vol. 19, no. 3, pp. 525--565, 2015.

\bibitem{ICWSM1510561}
M.~Cha, Y.~Gwon, and H.~T. Kung,
\newblock ``Twitter geolocation and regional classification via sparse
  coding,''
\newblock in {\em International AAAI Conference on Web and Social Media}, 2015,
  pp. 582--585.

\bibitem{LiuPaper}
J.~Liu and D.~Inkpen,
\newblock ``Estimating user location in social media with stacked denoising
  auto-encoders,''
\newblock in {\em Conference of the North American Chapter of the Association
  for Computational Linguistics {-} Human Language Technologies}, 2015, pp.
  201--210.

\bibitem{ICWSM136067}
D.~Jurgens,
\newblock ``That's what friends are for: Inferring location in online social
  media platforms based on social relationships,''
\newblock in {\em The International AAAI Conference on Weblogs and Social
  Media}, 2013, vol.~13, pp. 273--282.

\bibitem{compton2014geotagging}
R.~Compton, D.~Jurgens, and D.~Allen,
\newblock ``Geotagging one hundred million twitter accounts with total
  variation minimization,''
\newblock in {\em IEEE International Conference on Big Data}, 2014, pp.
  393--401.

\bibitem{RahimiCB17}
A.~Rahimi, T.~Cohn, and T.~Baldwin,
\newblock ``A neural model for user geolocation and lexical dialectology,''
\newblock in {\em Annual Meeting of the Association for Computational
  Linguistics}, 2017, pp. 209--216.

\bibitem{bengio2015deep}
Y.~Bengio, I.~J. Goodfellow, and A.~Courville,
\newblock ``Deep learning,''
\newblock {\em Nature}, vol. 521, pp. 436--444, 2015.

\bibitem{ZHAO201743}
J.~Zhao, X.~Xie, X.~Xu, and S.~Sun,
\newblock ``Multi-view learning overview: Recent progress and new challenges,''
\newblock {\em Information Fusion}, vol. 38, pp. 43--54, 2017.

\bibitem{krizhevsky2012imagenet}
A.~Krizhevsky, I.~Sutskever, and G.~E. Hinton,
\newblock ``Imagenet classification with deep convolutional neural networks,''
\newblock in {\em Advances in Neural Information Processing Systems}, 2012, pp.
  1097--1105.

\bibitem{zhou2017guided}
W.~Zhou, X.~Li, and D.~Reynolds,
\newblock ``Guided deep network for depth map super-resolution: How much can
  color help?,''
\newblock in {\em IEEE International Conference on Acoustics, Speech and Signal
  Processing}, 2017, pp. 1457--1461.

\bibitem{graves2013speech}
A.~Graves, A.~Mohamed, and G.~Hinton,
\newblock ``Speech recognition with deep recurrent neural networks,''
\newblock in {\em IEEE International Conference on Acoustics, Speech and Signal
  Processing}, 2013, pp. 6645--6649.

\bibitem{nguyen17}
D.~M. Nguyen, E.~Tsiligianni, and N.~Deligiannis,
\newblock ``Deep learning sparse ternary projections for compressed sensing of
  images,''
\newblock in {\em IEEE Global Conference on Signal and Information Processing
  [Available: arXiv:1708.08311]}, 2017.

\bibitem{zhang2012combining}
L.~Zhang, L.~Zhang, D.~Tao, and X.~Huang,
\newblock ``On combining multiple features for hyperspectral remote sensing
  image classification,''
\newblock {\em IEEE Transactions on Geoscience and Remote Sensing}, vol. 50,
  no. 3, pp. 879--893, 2012.

\bibitem{yu2012combining}
J.~Yu, D.~Liu, D.~Tao, and H.~S. Seah,
\newblock ``On combining multiple features for cartoon character retrieval and
  clip synthesis,''
\newblock {\em IEEE Transactions on Systems, Man, and Cybernetics, Part B},
  vol. 42, no. 5, pp. 1413--1427, 2012.

\bibitem{eisenstein2010}
J.~Eisenstein, B.~O'Connor, N.~A. Smith, and E.~P. Xing,
\newblock ``A latent variable model for geographic lexical variation,''
\newblock in {\em Conference on Empirical Methods in Natural Language
  Processing}, 2010, pp. 1277--1287.

\bibitem{priedhorsky2014inferring}
R.~Priedhorsky, A.~Culotta, and S.~Y.~D. Valle,
\newblock ``Inferring the origin locations of tweets with quantitative
  confidence,''
\newblock in {\em Conference on Computer supported Cooperative Work \& Social
  Computing}, 2014, pp. 1523--1536.

\bibitem{roller2012}
S.~Roller, M.~Speriosu, S.~Rallapalli, B.~Wing, and J.~Baldridge,
\newblock ``Supervised text-based geolocation using language models on an
  adaptive grid,''
\newblock in {\em Joint Conference on Empirical Methods in Natural Language
  Processing and Computational Natural Language Learning}, 2012, pp.
  1500--1510.

\bibitem{Wing_hierarchicaldiscriminative}
B.~Wing and J.~Baldridge,
\newblock ``Hierarchical discriminative classification for text-based
  geolocation.,''
\newblock in {\em Conference on Empirical Methods in Natural Language
  Processing}, 2014, pp. 336--348.

\bibitem{backstrom2010find}
L.~Backstrom, E.~Sun, and C.~Marlow,
\newblock ``Find me if you can: improving geographical prediction with social
  and spatial proximity,''
\newblock in {\em International Conference on World Wide Web}, 2010, pp.
  61--70.

\bibitem{journals/tgis/DavisPOA11}
C.~A.~Davis Jr, G.~L. Pappa, D.~R.~R. Oliveira, and F.~L. Arcanjo,
\newblock ``Inferring the location of twitter messages based on user
  relationships,''
\newblock {\em Transactions in GIS}, vol. 15, no. 6, pp. 735--751, 2011.

\bibitem{raghavan2007near}
U.~N. Raghavan, R.~Albert, and S.~Kumara,
\newblock ``Near linear time algorithm to detect community structures in
  large-scale networks,''
\newblock {\em Physical review E}, vol. 76, no. 3, pp. 036106, 2007.

\bibitem{RahimiCB15}
A.~Rahimi, T.~Cohn, and T.~Baldwin,
\newblock ``Twitter user geolocation using a unified text and network
  prediction model,''
\newblock in {\em Annual Meeting of the Association for Computational
  Linguistics and the International Joint Conference on Natural Language
  Processing}, 2015, pp. 630--636.

\bibitem{MassiveDataMining}
J.~Leskovec, A.~Rajaraman, and J.~D. Ullman,
\newblock {\em Mining of Massive Datasets},
\newblock Cambridge University Press, 2011.

\bibitem{DBLP:journals/corr/LeM14}
Q.~Le and T.~Mikolov,
\newblock ``Distributed representations of sentences and documents,''
\newblock in {\em International Conference on Machine Learning}, 2014, pp.
  1188--1196.

\bibitem{grover2016node2vec}
A.~Grover and J.~Leskovec,
\newblock ``node2vec: Scalable feature learning for networks,''
\newblock in {\em ACM SIGKDD International Conference on Knowledge Discovery
  and Data Mining}, 2016, pp. 855--864.

\bibitem{scikit-learn}
F.~Pedregosa, G.~Varoquaux, A.~Gramfort, V.~Michel, B.~Thirion, O.~Grisel,
  M.~Blondel, P.~Prettenhofer, R.~Weiss, V.~Dubourg, J.~Vanderplas, A.~Passos,
  D.~Cournapeau, M.~Brucher, M.~Perrot, and E.~Duchesnay,
\newblock ``Scikit-learn: Machine learning in {P}ython,''
\newblock {\em Journal of Machine Learning Research}, vol. 12, pp. 2825--2830,
  2011.

\bibitem{DBLP:journals/corr/MikolovSCCD13}
T.~Mikolov, I.~Sutskever, K.~Chen, G.~Corrado, and J.~Dean,
\newblock ``Distributed representations of words and phrases and their
  compositionality,''
\newblock in {\em Advances in neural information processing systems}, 2013, pp.
  3111--3119.

\bibitem{rehurek_lrec}
R.~Rehurek and P.~Sojka,
\newblock ``Software framework for topic modelling with large corpora,''
\newblock in {\em LREC 2010 Workshop on New Challenges for NLP Frameworks},
  2010.

\bibitem{twiter_geo_location_timing_matters}
M.~Dredze, M.~Osborne, and P.~Kambadur,
\newblock ``Geolocation for twitter: Timing matters.,''
\newblock in {\em Annual Conference of the North American Chapter of the
  Association for Computational Linguistics: Human Language Technologies},
  2016, pp. 1064--1069.

\bibitem{article:relu}
X.~Glorot, A.~Bordes, and Y.~Bengio,
\newblock ``Deep sparse rectifier neural networks,''
\newblock in {\em International Conference on Artificial Intelligence and
  Statistics}, 2011, pp. 315--323.

\bibitem{1984S.T.68R.158S}
R.~W. Sinnott,
\newblock ``Virtues of the haversine,''
\newblock {\em skytel}, vol. 68, pp. 158, Dec. 1984.

\bibitem{Bird:2009:NLP:1717171}
S.~Bird, E.~Klein, and E.~Loper,
\newblock {\em Natural Language Processing with Python},
\newblock O'Reilly Media, Inc., 1st edition, 2009.

\bibitem{DBLP:journals/corr/KingmaB14}
D.P. Kingma and J.~Ba,
\newblock ``Adam: A method for stochastic optimization,''
\newblock in {\em The International Conference on Learning Representations},
  2015.

\end{thebibliography}

\end{document}